# Conditional Utility, Utility Independence, and Utility Networks

**Yoav Shoham**[*]
Department of Computer Science
Stanford University
Stanford, CA 94306
shoham@cs.stanford.edu

## Abstract

We introduce a new interpretation of two related notions – conditional utility and utility independence. Unlike the traditional interpretation, the new interpretation render the notions the direct analogues of their probabilistic counterparts. To capture these notions formally, we appeal to the notion of *utility distribution*, introduced in previous paper. We show that utility distributions, which have a structure that is identical to that of probability distributions, can be viewed as a special case of an additive multiattribute utility functions, and show how this special case permits us to capture the novel senses of conditional utility and utility independence. Finally, we present the notion of *utility networks*, which do for utilities what Bayesian networks do for probabilities. Specifically, utility networks exploit the new interpretation of conditional utility and utility independence to compactly represent a utility distribution.

## 1 Introduction

There has recently been a growing interest within AI in representing and reasoning about utility. There are several reasons for this. First, while probabilistic methods have gained much influence, probability is only one ingredient of decision theory; foundations of decision theory are based on utility functions as much as they are on probability distributions. Second, just as there exist applications which call for reasoning purely about probabilities, there exist applications that call for reasoning purely about utilities. Examples include a software agent that needs to reason about the utility functions of other agents in a bargaining situation, and a meal-planning program needing to understand the gastronomic preferences of the user.

As we argue in previous paper [7][1], it would be quite convenient if we had a mechanism analogous to Bayesian networks to reason purely about utilities. As we further note there, at the heart of Bayesian networks lie three concepts: probability distribution, conditional probability, and probability independence. If we manage to mirror those notions in the case of utilities, we will have potentially availed ourselves of a ready-made mechanism for reasoning about utilities. In [7] we introduce the notion of utility distribution.[2] Here we concentrate on the notions of conditional utility and utility independence, and the derived notion of utility networks.

While not the main focus here, as we shall see, this paper does shed some new light on the notion of utility distribution itself. Specifically, while the treatment in [7] derives the notion from scratch, as a side effect of considering notions such as utility independence we will end up re-deriving the notion of utility distribution as an extension of standard decision theoretic notions, in particular those encountered in multiattribute utility theory (MAUT) [5, 3].

Indeed, most papers in AI that attempt to do something interesting with utilities appeal to MAUT, and to notions of conditionalization and independence therein. This is true of earlier work on influence diagrams that introduces multiple (additive or multiplicative) value nodes [6], and more recent papers by Bacchus and Grove [1] and by Doyle and Wellman [2].

I think that there are two reasons why researchers have concentrated on the classical notions of condi-tional utility and utility independence. First, the decision theoretic literature itself (most notably, Keeney and Raiffa's [5]) presents compelling arguments in favor of these notions. Second, the terms themselves suggest an analogy with conditional probability and probabilistic independence, leading to this vague hope that the utility-based notions will yield similar computational advantages.

---

[*] This work was supported in part by NSF grant IRI-9503109.

[1]Despite the publication dates, [7] describes work carried out almost a year before the work described here.

[2]We furthermore define the notion of a bi-distribution, which contains both a probability and utility distribution, but this will not play a major role in this paper.

The decision theory literature reinforces the analogy between the probabilistic and utilitarian notions. Here's a quote from [5]:

> One of the fundamental concepts of multiattribute utility theory is that of *utility independence*. Its role in multiattribute utility theory is similar to that of probabilistic independence in multivariate probability theory. (p.224)

I argue that this analogy is misleading. The classical sense of utility independence has fundamentally different properties from those of probabilistic independence. However, there exists a different sense of utility independence for which this analogy holds in a precise sense. The same is true of conditional utility.

Let me illustrate these two senses of conditional utility through an example. Referring to the hypothetical meal-planning program mentioned above, consider two conditional utilities:

- The utility for John of having beef for the main course, given that the appetizer is salmon mousse.

- The utility for John of having beef for the main course, given that John is vegetarian.

These are fundamentally different senses of conditional utility. The first conditions on an objective fact; loosely speaking, it can be thought of updating the utility based on information learned about the state of the world. The second conditions on a mental fact; loosely speaking, it can be thought of as updating the utility based on information learned about the preference structure of the agent. We might call the first 'objective' conditional utility, and the second 'subjective' conditional utility.

An analogy may be instructive here. The KR and database communities have learned to distinguish between updating a knowledge base and revising it [4]; the first reflects changes in the world, the second changes in information about the world. A similar distinction must be drawn with respect to conditional utilities.

The standard notion of conditional utility (and derived notions of utility independence) in decision theory is of the first variety, and it is the one commonly discussed in AI. However, this version of conditional utility is the one least similar to conditional probability. Perhaps for this reason, and despite great ingenuity on the part of the various authors, this notion has not yielded a computational device similar in nature and power to Bayesian networks.[3]

In the rest of this paper we do the following. First, we briefly review the basics of MAUT and the standard notion of utility independence (and conditional utility). Next we extend those notions to include the notion of utility distribution. We then formally present the alternative notions of utility independence and conditional utility, which are directly analogous to their probabilistic counterparts. We conclude with an a computational application of these ideas, and introduce the notion of utility networks.

## 2  Multiattribute utility theory (review)

Every utility function is defined over a set of states, and maps each state to a real value (its utility). In multi-attribute utility theory (MAUT) [3, 5] one posits a set of *n attributes* with corresponding domains $D_1, D_2, \ldots, D_n$, and the set of states is defined to be the Cartesian product $D_1 \times D_2 \times \ldots \times D_n$.

In general, specifying a MAU function can be expensive, exponential in the number of attributes. However, under special conditions the representation can be more compact. The general scheme for specifying these conditions goes like this: One defines certain "independence conditions" on the $n$ attributes, and then provides a "representation theorem," stating that under these independence conditions the utility function can be specified in a certain compact form. The remainder of this section summarizes these conditions and corresponding specialized representations.

We first note that a utility function $u$ over a set of states $S$ induces a preference ordering $\succeq_u$ on lotteries (or probability distributions) over $S$ via the expected utility construct:

$$p_1 \succeq p_2 \iff \Sigma_{s \in S} p_1(s) u(s) \geq \Sigma_{s \in S} p_2(s) u(s)$$

where $p_1$ and $p_2$ are any two lotteries over $S$. More generally, in the case of MAU functions, a utility func-

---

[3]Bacchus and Grove [1] do propose a graphical model in which to represent conditional independence, motivated by the analogy with Bayesian networks. However, their graphical model is very different from Bayesian networks (for example, arcs in it are undirected), and has not to date suggested a computational method. This is not detract from the contribution of this very interesting paper, only to correct possible misperception of its content.

tion defines preference on lotteries conditional on certain attribute $X$ having particular values $x$:

$$p_1 \succeq p_2(given X = x)$$
$$\Longleftrightarrow$$
$$\Sigma_{s \in S} p_1(s \mid X = x)u(s) \geq \Sigma_{s \in S} p_2(s \mid X = x)u(s)$$

Given this, we can define the general notion of utility independence. In the following definition, $Y$ and $Z$ are sets of attributes (in most applications, $Z$ will be the complement of $Y$):

**Definition 1 (based on [5], p.226)** $Y$ *is utility independent of $Z$ when conditional preferences on lotteries on $Y$ given $Z = z$ do not depend on the particular value of $z$.*

Using the notion of utility independence we define two independence conditions on a set of attributes. The first definition doesn't appear in the literature as a definition or given this name, but these will be convenient:

**Definition 2 (based on [5], p.293)** *Attributes $X_1, \ldots, X_n$ are singulary utility independent if every $X_i$ is utility independent of $\overline{X_i}$.*

The next definition involves stronger independence conditions:

**Definition 3 ([5], p.289)** *Attributes $X_1, \ldots, X_n$ are mutually utility independent if every subset of $\{X_1, \ldots, X_n\}$ is utility independent of its complement.*

The third and final type of independence involves a stronger condition than either of the first two:

**Definition 4 ([5], p.295)** *Attributes $X_1, \ldots, X_n$ are additive independent if preferences over lotteries on $X_1, \ldots, X_n$ depend only on their marginal probability distributions and not on their joint probability distribution.*

The relative strength of these three properties is reflected in the different representation theorems they allow. Starting off somewhat qualitatively, here is a rough description of four special forms of MAU functions that are based on these three independence conditions.

The first column specifies the independence condition on attributes, the second column names the special form in which the utility function can be represented, and the third gives some properties of this special form:

| ind. cond. | special form | some properties |
|---|---|---|
| singular | multilinear | $n$ simple utility functions; exponential number of constants; exponential number of additions and multiplications |
| mutual | *no standard name* | $n$ simple utility functions, $n$ constants, exponential number of additions and multiplications. However, this form always collapses to one of the following two special cases. |
| *none stated* | multiplicative | $n$ simple utility functions; $n$ constants; $n$ additions and multiplications. |
| additive | additive | $n$ simple utility functions; $n$ constants; $n$ additions and multiplications. |

Although the above table omits details, it does show that, as can be expected, stronger forms of utility independence admit more economical representations of the utility function. Indeed, additive independence (and even the more specific notion of *conditional* additive independence, in which additive independence holds given certain attributes) is the one that has attracted some interest recently in AI, and the one that is most relevant to us here. Given this, let us define the additive form of MAU functions precisely:

**Definition 5 (adapted from [5], p.295)**
*A MAU function $u$ over variables $x_1, \ldots, x_n$ is additive iff there exist functions $u_1, \ldots, u_n$ and constants $k_1, \ldots, k_n$ such that $u(x_1, \ldots, x_n) = \Sigma_{i=1}^n k_i u_i(x_i)$.*

(In fact, in the current context the constants $k_i$ play no role and could be omitted, but they will be useful in the sequel.)

## 3  From additive utility functions to utility distributions

One way in which to view our alternative perspective on utility independence is through a specialization of additive MAU functions. Our starting point is an example used by Bacchus and Grove. Consider two variables, $H$ (healthy) and $W$ (wealthy), and the following utility function:

$$u(HW) = 5, u(H\overline{W} = 2), u(\overline{H}W) = 1, u(\overline{HW}) = 0$$

It is easy to verify that $H$ is utility independent of $W$, and vice versa (intuitively, one prefers being healthy to being sick regardless of whether one is wealthy or poor, and vice versa). However, $H$ and $W$ are not additive independent. Consider two lotteries $p_1$ and $p_2$ defined by $p_1(HW) = 1/4, p_1(H\overline{W} = 1/4), p_1(\overline{H}W) = 1/4, p_1(\overline{H}\,\overline{W}) = 1/4$ and $p_2(HW) = 1/2, p_2(H\overline{W} = 0), p_2(\overline{H}W) = 0, p_2(\overline{H}\,\overline{W}) = 0$. $p_1$ and $p_2$ have identical marginals on $H$ and $W$, but while the expected utility of $p_1$ is 2, the expected utility of $p_2$ is $5/2$. Intuitively, a person with this utility function prefers the synergy of health and wealth more than is predicted by their individual utilities.

Now consider a modified example, in which $u(HW) = 3$ (and the other values remain unchanged). It is not hard to see that in this case $H$ and $W$ are additive independent.

This modified example is both instructive and misleading. It is instructive because it demonstrate that, unlike the notion of utility independence which is qualitative in nature (it merely compares various numbers), additive independence is arithmetic in nature. However, it is also misleading because this example has more properties than are required by the notion of additive independence. Specifically, the attributes in this example are binary in nature; you are either healthy or not. The more general case would allow for multiple values of health and of wealth.

However, while too specialized to be a neutral representative of additive utility functions, this example is representative of a new, more specific class, which we define next.

**Definition 6** *Given a vector of boolean attributes $X = x_1, \ldots, x_n$ (that is, each with domain $\{0, 1\}$), a utility function $u$ over $X$ is TIOLI ("take it or leave it") iff there exists constants $k_1, \ldots, k_n$ such that $u(x_1, \ldots, x_n) = \Sigma_{i=1}^n k_i x_i$.*

The interpretation of a TIOLI utility function is best explained through our modified example. Health contributes 2 to one's joy (or utility or satisfaction), Wealth contributes 1, and the total utility experienced in any given state is simply the sum of the joys supplied by the elements present in the state – $u(HW) = 2 + 1 = 3, u(H\overline{W}) = 2 + 0 = 2$, and so on. I will call the attributes of a TIOLI MAU function *utility factors* or simply *factors*, to denote the fact they are thought of as representing the basic ingredients of one's mental state of satisfaction.

It is worthwhile to mention here that MAUT is completely agnostic about the interpretation of attributes. In some examples each attribute is some good such as sugar or flour, its value denote the quantity of the good consumed in a state, and a state then becomes interpreted as a bundle of goods (on which one might bid in an auction, for example). In a different interpretation, the different attributes are days of the week, their values are the rewards experienced in each day, and their combination (for example, via addition or weighted addition) describes the overall utility experienced during the week. TIOLI functions admit the more psychological interpretation discussed above.

Finally, we note that the various $k_i$s can be translated and scaled to lie in the interval $[0..1]$ and to sum to 1. This yields the special form of TIOLI utility functions called *utility distributions*:

**Definition 7** *Given a vector of boolean attributes $X = x_1, \ldots, x_n$, a utility function $u$ over $X$ is a utility distribution iff there exists constants $k_1, \ldots, k_n$ such that (a) $0 \leq k_i \leq 1$, (b) $\Sigma_{i=1}^n k_i = 1$, and (c) $u(x_1, \ldots, x_n) = \Sigma_{i=1}^n k_i x_i$.*

Clearly, the structure of a utility distribution is essentially that of a probability distribution, except that the measure is applied to utility factors rather than to events. In fact, all that remains in order to make the two measures identical in structure is to lift the domain of the utility distribution to sets of factors. This is done in the obvious way; the utility of a set of factors is the sum of the individual utilities.[4]

The notion of utility distribution was introduced already in a previous paper [7], where it was defined independently of any pre-existing notion. I've re-derived the notion here in the context of multiattribute utility theory in order to be able to contrast different senses of, e.g., utility independence in the next section, but let me briefly mention here a few more elements discussed in [7].

The reader might be concerned about the applicability of these notions. We have discussed a progression of increasingly strong constraints on the structure of the utility function, and one might worry that the strongest constraint – the utility distribution form – is too rare to be of interest. It turns out that this is not so, at least in principle. For any utility function – even one not in MAU form – we can find a set of factors and a utility distribution over them, such that original utility function can be reconstructed from these factors. The only hitch is that the set of factors might not be as small as one would like. In the examples given above (the health/wealth, and the cars examples) the number of factors was logarithmic in the number of states. However, in general it will range between logarithmic and linear.

---

[4] We have seen that some standard special forms of MAU functions – in particular, the additive form – come with a representation theorem, stating necessary and sufficient conditions on the preference relation over lotteries that permit the special form in question. One might ask if similar necessary and sufficient conditions exist for TIOLI functions or utility distributions. This is a question that I haven't looked into closely thus far, but it seems that there do not exist simple, compact conditions. Obviously, a necessary condition is that the preference among lotteries depend only on the marginals, but it seems that the additional requirements needed to make this also a sufficient condition are not as neat. However, given my preliminary state of understanding, and the fact that this question will not play a role in the sequel, I won't pursue this further here.

I have not discussed here how one can combine probability distributions and utility distributions into one framework. In [7] I define the notion of a bi-distribution. Briefly, a bi-distribution is a structure consisting of a probability distribution and a utility distribution, with undirected arcs connecting some factors with some states. What is important to emphasize here is that one cannot in general define a probability distribution and a utility distribution on the same set. In effect, when one carves up the world into a set of elements, one usually makes a choice – these elements can be additive in probability (in which case they are called 'states') or additive in utility (in which case they are called 'factors'), but not both. However, given two such different sets of elements, one can define a third set by taking the Cartesian product of the first two. The elements in this induces set are additive in both probability and utility, but do not typically correspond to an intuitive concept.

## 4 Defining subjective senses of conditional utility and utility independence

Let us now reconsider the notions of conditional utility and utility independence, in the context of utility distributions. Since utility distributions are MAU functions, we can apply (any of) the standard notions of conditioning and independence to them. However, unlike arbitrary MAU functions, utility distributions also allow us to define the subjective versions of these two notions, discussed in the introduction.

Let's start with conditional utility. In the subjective version of the notion, we interpret $u(x|y)$ as "the utility of $x$, given that $u(y) = 1$, or, that all the utility is derived from the factor set $y$."

This can be explained intuitively through an example. Consider a person whose entire value system is based on owning any or all of three cars, a Rolls Royce, a Maserati, and a Ford. We define three corresponding utility factors, with the $k_i$s defined by $u(r) = 0.1$,

$u(m) = 0.2$, and $u(f) = 0.7$. Thus, for example, if the person owned all three cars he would derive 0.7 of his utility from the Ford; this can be thought of as the prior utility of a Ford. But what would the contribution of the Ford to utility be if it is learned that the person derives no pleasure from British-made cars? By direct analogy with probabilities, we define subjective conditional utility by $u(x|y) = u(x \cap y)/u(y)$. This gives us, in particular, that the utility of a Ford for a British-car-hater is

$$u(f|fm) = u(f \cap fm)/u(fm) = u(f)/u(fm) = 0.7/(0.7 + 0.2) = 7.77\ldots$$

Subjective utility independence will also be defined similarly; factor set $x$ will be said to be utility-independent of $y$ iff $u(x|y) = u(x)$. The intuition behind this property will be that the relative contribution of the factor set $x$ does not change if we learn that the entire contribution to joy lies within the set $y$. Thus in the car example, a Ford is not utility independent of non-British cars, since

$$u(f) = 0.7 \neq 0.77\ldots = u(f|fm)$$

However, if we add a Toyota and modify the $k_i$s as follows:

| factor | r | m | f | t |
|--------|------|------|-------|------|
| utility | 6/30 | 3/30 | 14/30 | 7/30 |

then it's easy to verify that the utility of the class of cars made in English-speaking countries is independent of the utility of European-made cars:

$$u(fr|rm) = u(fr) = 0.66\ldots$$

## 5 Utility networks

Our goal at the outset was to investigate the possibility of endowing utilities with the properties of probabilities, so that the benefit of Bayesian networks can transfer to them. We've now achieved this, so it might be argued that the rest of the story is anti-climactic. Since now we have notions of utility distribution, conditional utility, and utility independence that have exactly the mathematical properties of their probabilistic counterparts, we can, it might be argued, go ahead and use Bayesian networks to represent and reason about utilities. (Of course, the term Bayesian networks should now be replaced by something more appropriate. We might use terms *utility network*, or *u-net*, when we are using the Bayesian-network-like structure for utilities, and *p-net* in the case of probabilities.)

Left at this, however, this might be deemed little more than a formal exercise. There are at least two sources of potential complaint. The first is that utility factors might not correspond to anything intuitive, and hence could not be used in practice. The second is that Bayesian networks have proven useful because the structure of such networks reflects *causality*; it is our intuitive grasp of causal relations in the world that allows us to construct and understand Bayesian networks. What's to help us make intuitive sense of utility networks? Let me address these two potential complaints in turn.

I don't know whether utility factors will in general turn out to be intuitive or not; I think we don't have enough experience to pronounce judgement on this. However, we do not need to reason about factors directly. This is exactly analogous to probabilistic reasoning. Bayesian networks do not represent individual world states; in any realistic domain, these would be impossibly complex for any human to comprehend. Rather, each variable represents an event, a set of states, which abstracts away from all but a few aspects of the world. The events "it rained," "the lawn was watered," and "the pavement is wet," are examples of such abstractions. We will do the same with factors, and reason only about sets of factors, such as "having money," "being admired by a loved one," and "owning a motorcycle."

If these sets of factors seem indistinguishable from events, this is no accident. While in general the set of individual states and the set of individual factors are disjoint, certain *sets* of states and *sets* of factors might be co-extensive. In [7] I discuss how for any utility function defined on states one can construct a dual, factor space such that the utility of each original state is the sum of some subset of factors in the dual space. In general, every event (set of states) defines a set of factors, and every set of factors defines an event. When a given set of factors and a given event define each other, they are co-extensive.

Detailed discussion of this is beyond the scope of the paper. Hopefully the examples to come will give more intuition, and in particular clarify the general point about not having reason about individual factors. But here's the most important point. We shouldn't fret over whether nodes in a network are events, sets of factors, or both. The critical question is the interpretation we give to the links in the network, that is, to the relationships between the nodes.

Which brings us to the second issue, causality. Although I'm in general suspicious about cavalier uses of the causal terms in AI, and in particular about the purely causal interpretation of Bayesian networks, there is no denying that most of the Bayesian networks actually produced in fact correspond to intuitive no-

tions of causality, typified by the standard example:

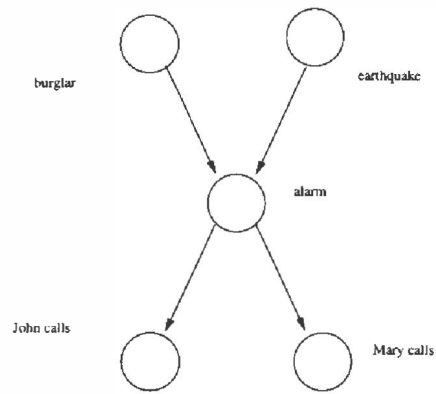

The intuitive interpretation of this example is that either an earthquake or a burglary can trigger your house alarm, which in turn can cause both of your neighbors – John and Mary – to call you at work. Indeed, it is hard to imagine how one would come up with this network without appealing to causality. Similarly, given the causal interpretation, it is intuitively clear why John's calling you is probablistically independent of there being an earthquake given that the house alarm went off.

Can any concept play the role of causality in utility networks? I think the answer is yes, and that the concept is *teleology*, a form of "mental causality." Consider two factor sets, "love of art" and "visiting the San Francisco Museum of Modern Art (SFMOMA)." The utility one places on Art determines, at least in part, the utility of SFMOMA. In other words, SFMOMA is desired inasmuch as it contributes to satisfying the Art desire. It seems to me that this familiar utilitarian notion serves just the purpose needed here. Consider a more elaborate example:

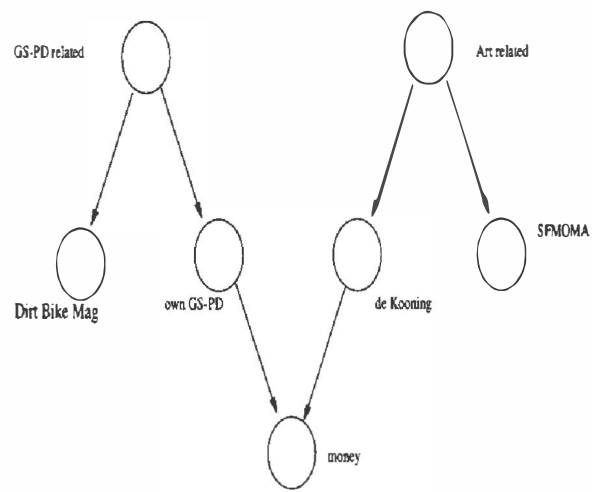

The way to interpret this picture is as follows. The person whose utility is modeled has two basic moti-vators – love of BMW GS-PD motorcycles, and love of art. In service of the GS-PD motivator, the per-son place a certain value on owning one of these bikes, and on reading Dirt Bike magazine, which covers dual-purpose motorcycles such as the GS-PD. The Art mo-tivator leads him to desire to go to SFMOMA, as well as to own an original de Kooning. Both the consider-ation of owning a GS-PD and that of owning the de Kooning lead the person to place a certain value on money. The graph structure induces various indepen-dence conditions. For example, the utility of money given that the person wishes to purchase a GS-PD is independent of his love of motorcycles.

This is presumably a natural and familiar pattern. In-deed, it is a causal pattern, but one must be care-ful about the nature of this causal relation. Consider again the link between Art and SFMOMA. There are at least three causal connections one might be tempted to identify here. First, satisfying the SFMOMA desire will cause the higher satisfaction of the desire for Art. Second, the desire for Art will cause the person to de-sire to visit SFMOMA. Third, the desire for Art will cause the person to actually visit SFMOMA. *In utility networks, the links capture the first two kinds of causal-ity, but not the third.* Utility networks do not speak about what will be the case, only about a person's mental state. Indeed, whether the person will actually visit the museum might be determined by factors that are unrelated to the person's preference structure, just as even the most intense interest in the GS-PD will not necessarily result in the person buying one.

This is not to say that mental state doesn't impact re-ality, only that utility network don't capture this fact. Although this lies beyond the scope of this paper, let me add a few words on reasoning simultaneously about probabilities and utilities, in a structure we might call a *bi-network*. I've mentioned already that in [7] I de-fine a bi-distribution, which couples a probability dis-tribution (or p-distribution) with a utility distribution (or u-distribution). We can represent a bi-distribution by a pair of networks, a Bayesian network (or p-net) and a utility network (or u-net). If the two were un-related that wouldn't be interesting, but in fact a bi-distribution includes a set of undirected edges between nodes in the two distributions, which can be used to induce utilities on the p-net and probabilities on the u-net.

In a simple version of bi-networks, computation in each net will proceed independently; in particular, we can condition the two nets independently from one an-other. Just as influence diagrams use utility nodes merely to compute values resulting from the proba-bilistic conditioning, in simple bi-networks links will be used to merely compute expected-utility values re-sulting from both probabilistic and utilitarian condi-tioning.

A more ambitious version of bi-networks will allow "hybrid" forms of conditioning, in which the p-net and the u-net share nodes, and, more interestingly, one can condition probabilities on utilities and vice versa. However, this remains an avenue for future investiga-tion.

## 6  Summary

The ideas described in this paper are part of a con-tinuing enquiry into the role of choice theory in AI, and the questioning of some established assumptions in choice theory. The main messages synthesized so far as a result of this inquiry are as follows:

- There is no reason for the traditional asymmetries between probabilities and utilities. In particular, utilities too can enjoy distributions. This is the main focus of [7], and was discussed here only partially,

- There exists a sense of conditional utility that is different from the classical one, and utility dis-tributions provide a way to define it. The same is true of utility independence. This has been the primary focus of this paper. For this reason, the related work discussed throughout the paper is mostly that which pertains to utility indepen-dence.

- The interpretation of classical results in decision theory, and in particular the von Neumann and Morgenstern representation theorem, is opened to question. This is discussed in [7] but not here.

- Utility networks can do for utilities what Bayesian networks do for probabilities, with the concept of teleology replacing that of causality. This was discussed in the previous section.

- The new perspective suggests a structure, called a bi-distribution, in which probability and util-ity distributions live side by side, and which can be used to compute expected utilities. This too is discussed in [7] but only mentioned here, with brief discussion of how it suggests the notion of bi-networks as a generalization of both Bayesian networks and utility networks.

**Acknowledgements.** I have discussed my ideas on utilities with many people. For the material in this article, I thank in particular participants in the 1997 AAAI Spring Symposium on Qualitative Decision Theory, Adam Grove, and several anonymous and very careful referees. Which is not to suggest that any of the above endorse the ideas expressed herein.